\documentclass[12pt]{article}
\usepackage{a4}
\usepackage{epsf}

\newcommand{\id}{\mbox{1$\!\!$I}}
\newcommand{\be}{\begin{equation}}
\newcommand{\ee}{\end{equation}}
\newcommand{\bea}{\begin{eqnarray}}
\newcommand{\eea}{\end{eqnarray}}

\newcommand{\nn}{\nonumber}

\newcommand{\dedouble}{ \stackrel{ \leftrightarrow }{ \partial } }

\begin{document}

\begin{center}
\title{Quenched Supersymmetry}
\author{ A. Donini$^a$, E. Gabrielli$^b$, 
M.B. Gavela$^c$} \par \maketitle
Departamento de F\'\i sica Te\'orica, 
Universidad Aut\'onoma de Madrid, Canto Blanco, 28049 Madrid\footnote{ 
E-mail: $^a$ donini@daniel.ft.uam.es, $^b$ egabriel@daniel.ft.uam.es,
$^c$ gavela@garuda.ft.uam.es.}. \\
\end{center}

\begin{abstract}

We study the effects of quenching in Super-Yang-Mills theory.
While supersymmetry is broken, the lagrangian acquires a new flavour
$U(1 \mid 1)$ symmetry. The anomaly structure thus differs from the
unquenched case.
We derive the corresponding low-energy effective lagrangian. 
As a consequence, we predict the mass splitting expected in numerical
simulations for particles belonging to the lowest-lying supermultiplet.
An estimate of the systematic error due to quenching follows.

\end{abstract}
\begin{flushright} FTUAM/98/12\\  hep-th/9810127
\end{flushright}
\newpage

\section{Introduction}
\label{intro}

If it finally turns out that nature can express itself in the 
language of supersymmetric theories, the understanding of the strongly 
coupled regime of the latter may be very important. From gluino condensation
to many other issues in nowadays implementations of string theories,
such understanding is pertinent. A first important step in this direction
was made by Veneziano and Yankielowicz (VY) \cite{vy}, when they derived the
low energy effective lagrangian for pure $N=1$ supersymmetric Yang-Mills 
theory (SYM). Recently, several modifications to this lagrangian have been 
proposed \cite{ks,fgs}. 

The ``a priori'' primary tool for a direct study of strongly coupled field
theories is lattice regularization. Even though the lattice regulator by itself 
explicitly breaks supersymmetry and chiral symmetry, it is possible to implement a 
suitable procedure to recover the SUSY and anomalous chiral Ward identities
in the continuum limit \cite{cv}.
Some numerical results for $N=1$ SYM can be found in \cite{m1,dg,desy}. 
However, the numerical implementation of the full theory
can be very time/resource consuming due to the fermion determinant
computation\footnote{In the case of SYM, actually, we deal with an object
even harder to handle numerically than the determinant, the Pfaffian.}.
It is well known that the quenched approximation greatly reduces such costs, 
albeit at the price of the corresponding systematic error. 
Although supersymmetry is broken upon quenching, those simulations represent a
convenient first step to explore the parameter space of the theory, as far 
as the induced error is under control, one of the issues adressed here.

A qualitative and quantitative understanding of the effects of quenching 
in a supersymmetric theory should come out of the theoretical endeavour 
of the present work. In this paper we derive the low energy effective lagrangian 
for quenched $N=1$ SYM theory in the continuum, 
paralleling the work done for QCD in \cite{bg}. 
Our work is based on the VY effective lagrangian (modified theories
could be studied within the same approach). We focus in the resulting 
mass spectrum for the low-energy supermultiplet and the mass splitting induced
by quenching. A preliminary version of our results can be found in \cite{dgg}. 

The paper is organized as follows: in Sect.~\ref{symsec} we recall the 
$N=1$ SYM theory and its anomaly structure, as well as 
its low-energy effective theory; in Sect.~\ref{qsymsec} we investigate  
quenched SYM and its anomalies, whereas in Sect.~\ref{qVYsec} 
the low-energy effective quenched lagrangian is proposed. Sect.~\ref{secmass}
deals with the analytical computation of  the mass spectrum 
for the quenched extension of the lowest-lying VY supermultiplet and
in Sect.~\ref{concl}, we eventually draw our conclusions.

\section{Super-Yang--Mills action and its symmetries}
\label{symsec}

Consider the N=1 Super-Yang-Mills lagrangian, which is the minimal
supersymmetric version of a pure $SU(N_c)$ gauge theory. It describes a 
vector supermultiplet with fermion and boson fields
in the adjoint representation of the colour gauge group $SU(N_c)$. 
No matter superfields,
containing fermion and scalar fields in the fundamental representation,
are present. The on-shell action is given by
\be
\label{symaction}
S_{SYM} = \int d^4 x \left \{ - \frac{1}{4} F^a_{\mu\nu} F^{a\mu\nu}
 + \frac{i}{2} \bar \lambda^a \gamma^\mu D^{ab}_\mu \lambda^b \right \} ,
\ee
where $a,b= 1,\dots ,(N_c^2-1)$ are indices running in the adjoint 
representation of $SU(N_c)$ and $D^{ab}$ is the corresponding 
covariant derivative acting on the gluino field $\lambda^a$, 
which is a Majorana fermion. 
At the classical level, the action is also invariant under a 
chiral $U(1)$ symmetry acting on the gluino field\footnote{
Due to the Majorana nature of gluinos there is no vector 
$U(1)$ symmetry.}, as well as scale invariant. 
At the quantum level both these symmetries are broken, though,
by the corresponding chiral and trace anomalies,
\be
\left . \begin{array}{ccc}
\partial^\mu J_\mu &=& - c(g) F^a_{\mu\nu} 
\tilde F^{a\mu\nu} \, , \\
\Theta^\mu_\mu &=& c(g) F^a_{\mu\nu} F^{a\mu\nu} \,.
\end{array} \right .
\label{anomalies}
\ee
$J_{\mu}$ and $\Theta_{\mu}^{ \mu}$ denote the 
chiral current and the trace of the energy momentum tensor $\Theta_{\mu \nu}$,
respectively, while $c(g) = \beta(g)/2 g$,
with $\beta(g)$ being the one-loop $\beta$-function of the theory.
The two anomalies in eq.~(\ref{anomalies}) 
and the supersymmetric trace anomaly, given by 
$\gamma^\mu S_\mu = 2 c(g) \sigma_{\mu\nu} F^a_{\mu\nu} \lambda^a$,
 belong to the same supermultiplet.

From the assumption of colour confinement in SYM, for which some supporting
numerical results exist \cite{tro},
it follows that the spectrum should contain colourless bound states 
of gluons and gluinos. In the limit  $N_c\to \infty$ the low energy
spectrum will contain both mesons and baryons, made up of an even
and odd number of gluinos, respectively (while in QCD only mesons
survived as light fields). 
Moreover, these bound states should be described by a supersymmetric low energy 
effective theory which has to reproduce all the symmetries 
of the fundamental action including its chiral, scale and supersymmetric
anomalies.
Such an action, here below dubbed $S_{VY}$, was derived by VY in \cite{vy}. 
They considered the lowest dimensional 
composite operators (made of gluons and gluinos)
belonging to the same chiral supermultiplet, with the result
\be
S_{VY} = \int d^4 x \left \{ \frac{9}{\alpha} (S^\dagger S)^{1/3}_D + 
 \left [ \frac{1}{3} \left ( S \log (\frac{S}{\mu^3}) - S \right )_F + h.c. 
\right ] \right \} ,
\ee
where $S$ is a chiral supermultiplet containing the bound states 
and $\alpha$ and $\mu $ are free parameters.
Notice that the request to reproduce 
the correct anomalies of the fundamental action fixes completely the
form of the superpotential.

Expanding $S$ in component fields, the off-shell VY action can be written as
\bea
S_{VY} &=& \int d^4 x \left \{ 
\frac{1}{\alpha (\phi^\star \phi )^{2/3}} 
              \left [              
  \partial_\mu \phi^\star \partial^\mu \phi 
       + i \bar \chi \gamma^\mu \partial_\mu \chi \right ] \right . \nn \\
&+& \left . \frac{1}{\alpha (\phi^\star \phi )^{2/3}} 
\left [ \frac{4}{9} \frac{(\bar \chi_R \chi_L) (\bar \chi_L \chi_R)}{\phi^\star \phi}
       - \frac{2}{3} \left ( M \frac{\bar \chi_L \chi_R}{\phi^\star} +
                             M^\dagger \frac{\bar \chi_R \chi_L}{\phi}  \right ) + M^\dagger M
        \right ] \right . \nonumber \\
&-& \left . \frac{1}{3 \alpha} \frac{(\bar \chi \gamma^\mu \gamma^5 \chi) 
( \phi^\star \dedouble_\mu \phi)}{ (\phi^\star \phi)^{5/3} } \right . \nonumber \\
&-& \left .
\frac{1}{3} \left ( \frac{\bar \chi_R \chi_L}{\phi} + \frac{\bar \chi_L \chi_R}{\phi^\star}
                  \right )
- \frac{1}{3} \left [ M \log (\frac{\phi}{\mu^3}) +
 M^\dagger \log (\frac{\phi^\star}{\mu^3}) \right ] \right \}, 
\label{offVY}
\eea
where $\phi$ is a complex scalar field, $\chi$ is a Majorana spinor, 
$\chi_{R,L} = \frac{1}{2} ( \id \pm i \gamma_5) \chi $
and $M$ is a complex auxiliary field\footnote{We get a
-1/2 factor in front of the $\bar \chi \gamma_\mu \gamma_5 \chi$ term
with respect to eq. (22) of \cite{vy}. We have explicitly checked that
our result satisfies supersymmetry.}.
In terms of the fundamental fields, they are described by
\be
\label{fields}
\left . \begin{array}{ccc}
\phi &=& c(g) \bar \lambda^a_R \lambda^a_L \,,\\
\chi &=& \frac{i c(g) }{2} \sigma_{\mu\nu} F^{a\mu\nu} \lambda^a \,,\\
   M &=& - \frac{c(g)}{2} \left ( F^a_{\mu\nu} F^{a\mu\nu} 
                           + i F^a_{\mu\nu} \tilde F^{a\mu\nu} \right )\,, 
\end{array} \right .
\ee
where $c(g)$ is the same factor appearing in the anomalies,  
eq.~(\ref{anomalies}).
Observe that in QCD the $F^a_{\mu\nu} \tilde F^{a\mu\nu}$ field 
can be regarded as an auxiliary field only in the low-energy and
$N_c\rightarrow \infty$ limits. Here, due to supersymmetry, 
$F^a_{\mu\nu} F^{a\mu\nu} $  and $F^a_{\mu\nu} \tilde F^{a\mu\nu}$
have no kinetic terms and appear at most quadratically in the lagrangian, 
eq.~(\ref{offVY}): they automatically play the role of auxiliary fields.

In order to obtain canonical kinetic terms, $\phi$ and $\chi$ can be 
rescaled as follows:
\be
\left . \begin{array}{ccc}
\phi & \rightarrow \left ( \frac{\alpha^{3/2}}{27} \right ) \phi^3\,, \\
\\
\chi & \rightarrow \left ( \frac{\alpha^{3/2} }{9 \sqrt{2}} \right ) 
                                  \phi^2 \chi \, ,
\end{array} \right .
\ee
allowing to rewrite eq.~(\ref{offVY}) as
\bea
S_{VY} &=& \int d^4 x \left \{ 
  \partial_\mu \phi^\star \partial^\mu \phi 
       + \frac{i}{2} \bar \chi \gamma^\mu \partial_\mu \chi  \right . \nn \\
&+& \left . 
 \frac{(\bar \chi_R \chi_L) (\bar \chi_L \chi_R)}{\phi^\star \phi}
       - \frac{9}{\alpha^{3/2}} \frac{1}{(\phi^\star \phi)}
\left ( M \frac{\bar \chi_L \chi_R}{\phi} +
        M^\dagger \frac{\bar \chi_R \chi_L}{\phi^\star} \right ) 
        \right . \nonumber \\
&+& \left . \frac{81}{\alpha^3 } \frac{1}{(\phi^\star \phi)^2} M^\dagger M
- \frac{1}{2} \frac{(\bar \chi \gamma^\mu \gamma^5 \chi) 
( \phi^\star \dedouble_\mu \phi)}{ (\phi^\star \phi) } \right . \nonumber \\
&-& \left . \frac{\alpha^{3/2}}{18} 
\left ( \bar \chi_R \chi_L \phi + \bar \chi_L \chi_R \phi^\star \right )
- \left [
M \log (\frac{\sqrt{\alpha} \phi}{3 \mu}) + M^\dagger 
\log (\frac{\sqrt{\alpha} \phi^\star}{3 \mu})
\right ] \right \}. 
\label{offVYcan}
\eea
 The use of the equation of motion for the auxiliary field $M$,
\be
M = \frac{\alpha^3}{81} (\phi^\star \phi)^2 
\log (\frac{\sqrt{\alpha}\phi^\star }{3 \mu})
     + \frac{\alpha^{3/2}}{9} ( \bar \chi_R \chi_L \phi ),
\ee
eventually leads from eq. (\ref{offVYcan}) to the on-shell VY action,
\bea
S_{VY} &=& \int d^4 x \left \{ 
  \partial_\mu \phi^\star \partial^\mu \phi 
       + \frac{i}{2} \bar \chi \gamma^\mu \partial_\mu \chi \right . \nn \\
&-& \left . \frac{1}{2} \frac{(\bar \chi \gamma^\mu \gamma^5 \chi) 
( \phi^\star \dedouble_\mu \phi)}{ (\phi^\star \phi) }
- \frac{\alpha^{3/2}}{18} 
\left ( \bar \chi_R \chi_L \phi + \bar \chi_L \chi_R \phi^\star \right ) 
 \right . \nn \\
&-& \left . \frac{\alpha^{3/2}}{9} \left [ 
  \bar \chi_R \chi_L \phi \log (\frac{\sqrt{\alpha} \phi}{3 \mu}) 
+ \bar \chi_L \chi_R \phi^\star \log (\frac{\sqrt{\alpha} \phi^\star}{3 \mu})
\right ] \right . \nn \\
&-& \left . \frac{\alpha^3}{81} (\phi^\star \phi)^2 
\log (\frac{\sqrt{\alpha} \phi^\star}{3 \mu})
\log (\frac{\sqrt{\alpha} \phi}{3 \mu})  \right \}\,. 
\label{onVYcan}
\eea
Using an exponential representation for the complex scalar field,
$\phi\equiv \rho_{S} e^{i \theta_S}/\sqrt{2} $, where $\rho_{S}$ 
and $\theta_S$ are real scalar and pseudoscalar fields, respectively, 
the VY scalar potential $V_{VY}$ can be written as
\be
\label{V_VY}
V_{VY} =\frac{\alpha^3}{81} \frac{\rho_S^4}{4} \left[
\log^2 (\frac{\sqrt{\alpha}}{3\sqrt{2}} \frac{\rho_{S}}{ \mu})+
\theta_S^2 \right].
\ee
The minimum of the potential is found at a non-zero value of $\rho_S$,
and therefore spontaneous chiral symmetry breaking occurs. Nevertheless, 
the would-be goldstone boson, $\theta_S$, is not a massless field: 
the anomaly terms in the lagrangian explicitly break the symmetry,
providing a mass scale. As supersymmetry is unbroken, 
mass degeneracy among the members of the multiplet is 
preserved with the result $m_\theta\,=\,m_\rho\,=\,
m_\chi\,=\,\frac{1}{3} \alpha \mu$.

\section{Quenched Super-Yang--Mills and its symmetries}
\label{qsymsec}

This section describes how to implement the quenched approximation 
in N=1 SYM theory and discusses which new symmetries appear as a 
consequence of the quenching procedure.

Alike to the approach in ref.~\cite{bg}, we add to
the SYM lagrangian a ghost scalar field in the adjoint representation. 
This new field,  $\eta^a$, has the same quantum numbers as the gluino
field, $\lambda^a$, although ``wrong'' spin-statistics.
 Containing a ghost-like field, the new lagrangian violates unitarity,
 not a surprise as the quenched approximation automatically does so. 
We thus propose the following quenched fundamental action:
\be
\label{qsymaction1}
S^q_{SYM} = \int d^4 x \left \{ - \frac{1}{4} F^a_{\mu\nu} F^{a\mu\nu}
+  \frac{i}{2} \bar \lambda^a \gamma^\mu D^{ab}_\mu \lambda^b 
 + \frac{i}{2} \bar \eta^a ( i \gamma^\mu \gamma_5 ) D^{ab}_\mu \eta^b 
 \right \}.
\ee
It should be noticed that the kinetic term of the new field is no longer 
of the usual $\gamma^\mu D^{ab}_\mu$ fermionic type, as 
$\bar \eta^a \gamma^\mu D^{ab}_\mu \eta^b = 0$ (up to total derivatives)
due to the ``wrong'' spin-statistics behaviour of the ghost,
in the same way as $\bar \lambda^a \gamma^\mu \gamma_5 D^{ab}_\mu \lambda^b = 0$. 
The new $\gamma^\mu \gamma_5 D^{ab}_\mu$ kinetic operator 
is thus the only possible kinetic term for these 
Majorana ghosts. This situation differs from that in ref.~\cite{bg}, 
as quenching QCD requires the inclusion of Dirac ghosts, 
for which the canonical fermionic kinetic term does exist. 

The fermionic integration over the Majorana fields in the 
generating functional of SYM gives the Pfaffian operator $Pf(O)$
(where $O$ is the Dirac operator). The usual fermion determinant, $\det(O)$,
is defined to be its square. In order to implement the quenched approximation
the internal fermion loops, represented by the Pfaffian, should be canceled.
Our new kinetic term in eq.~(\ref{qsymaction1}) allows such a 
cancelation, since $\det (O \gamma_5 ) = \det (O) \det (\gamma_5) = \det (O)$ in $D=4$.
After the integration over the ghost degree of freedom in the 
generating functional, the ghost and fermion Pfaffians cancel each other,
thus implementing the quenched approximation\footnote{
Some problems could appear in perturbation theory when
using dimensional regularization. It is probably necessary to use
the dimensional reduction scheme for the fermion-ghost loop
cancellation to work.}. 

The action $S^q_{SYM}$ is certainly no longer supersymmetric, as new 
bosonic fields have been introduced with no fermionic counterparts. 
It is still gauge invariant and classically scale invariant
and its $U(1)$ chiral symmetry is promoted to a
$U(1 \mid 1)$ chiral symmetry, the latter corresponding to the invariance 
of the lagrangian under the exchange of the fermion with the ghost. 
Since for $N_f = 1$ Majorana fermion there is no $U(1)$ fermion number
symmetry, no vector $U(1 \mid 1)$ symmetry is expected.

The $U(1 \mid 1)$ group is a $Z_2$ graded Lie group with both bosonic and 
fermionic generators (the supersymmetric algebra itself obeys a $Z_2$ graded Lie group). 
The associated algebra is defined by 
two bosonic generators $\left\{\sigma_0, \sigma_3\right\}$ 
and two fermionic generators $\{\sigma_1, \sigma_2 \}$,
where $\sigma_0$ is the identity matrix, and $\sigma_{1,2,3}$ denote the 
Pauli matrices. The graded algebra consists of the following 
commutation and anticommutation rules 
\be
\label{z2algebra}
\left . \begin{array}{ccc}
 \left [ \sigma_0, \sigma_i \right ]   &=& 0\,, \\
 \left [ \sigma_1, \sigma_3 \right ]   &=& - \sigma_2\,,  \\
 \left [ \sigma_2, \sigma_3 \right ]   &=&   \sigma_1 \,, \\
 \{ \sigma_i , \sigma_j \} &=& 2 \delta_{ij} \sigma_0 \,.
\end{array} \right .
\ee

Any element of the $U(1 \mid 1)$ group is a $2 \times 2$ matrix with
diagonal bosonic elements and off-diagonal fermionic ones.
The group is unitary in the usual sense $U^\dagger U = I$.
As for ordinary Lie groups, any element of $U(1\mid 1)$ can
be represented as a continuous function of four parameters $\epsilon_i$ as
follows
\be 
U(\epsilon) = \exp{ \left\{ i \sum_{i=0}^{3} \epsilon_{i} \sigma_{i}
\right\}}\,,
\ee
where $\epsilon_{0,3}$ are real numbers and $\epsilon_{1,2}$ 
are real Grassman numbers.

Invariants under the $U(1 \mid 1)$ group can be constructed using the 
cyclic properties of the superstrace $Str$, defined as
\be
Str \left ( \begin{array}{cc} a & b \\ c & d \end{array} \right ) 
= a - d \,,
\ee
where, in general, $a,d$ are complex numbers and $b,c$ complex Grassman
numbers. An extended description of the properties of  
graded groups can be found for instance in \cite{dewitt}.

It is possible to rewrite the action in eq.~(\ref{qsymaction1}) 
in a more compact form,
\be
\label{qsymaction2}
S^q_{SYM} = \int d^4 x \left \{ - \frac{1}{4} F^a_{\mu\nu} F^{a\mu\nu}
 + i \bar Q^a_R \gamma^\mu D^{ab}_\mu Q^b_R \right \}\, , 
\ee
where $Q$ is the doublet $Q^a=( \lambda^a , \eta^a )$.
Written so, the lagrangian is now manifestly invariant under 
chiral $U(1 \mid 1)$ transformations, defined as follows:
\be
\label{U_11trans}
\left . \begin{array}{ccc}
     Q_R & \rightarrow U Q_R 
& =  \exp{\left\{  i \frac{\alpha_i \sigma^i}{2} \right\}} Q_R \,,\\
Q_L & \rightarrow  U^\dagger Q_L \,. &
\end{array} \right .
\ee
The four currents associated to the axial $U(1\mid 1)$ symmetry are
\be
J^i_\mu = \bar Q^a_R \sigma^i \gamma^\mu Q^a_R 
\ee
(where $i = 0, \dots, 3$), or, explicitly,
\be
\label{z2currents}
\left . \begin{array}{ccc}
J_\mu^0 &=&\frac{1}{2} ( i \bar \lambda^a \gamma_\mu \gamma_5 \lambda^a + 
\bar \eta^a \gamma_\mu \eta^a )\,,\\ 
J_\mu^+ &=& \bar \lambda^a_R \gamma_\mu \eta^a_R \,,\\ 
J_\mu^- &=& \bar \eta^a_R \gamma_\mu \lambda^a_R \,,\\ 
J_\mu^3 &=& \frac{1}{2} (i\bar \lambda^a \gamma_\mu \gamma_5 \lambda^a - 
\bar \eta^a \gamma_\mu \eta^a )\,. 
\end{array} \right .
\ee
Recall that the currents transform under $U( 1 \mid 1)$ as 
$(\sigma^i J^i_\mu) \to U (\sigma^i J^i_\mu) U^\dagger$. 
The fermionic currents $J_\mu^{\pm}$, which mix fermions and ghosts,
cannot give rise to anomalies. To identify which new currents 
in eq.~(\ref{z2currents}) are anomalous, let us consider the 
triangle graph for the ghost field. 
Define the triangle graph $\Gamma_{\mu \nu \rho}$ as 
\be
\label{triang}
\left . \begin{array}{ccc}
\Gamma^{\lambda}_{\mu \nu \rho} &=& 
\langle 0| T \{ A_{\mu}^{\lambda}(x) 
V_{\nu}^{\lambda}(y) V_{\rho}^{\lambda}(0) \} |0 \rangle \,,\\
\Gamma^{\eta}_{\mu \nu \rho} &=& 
\langle 0| T \{ A_{\mu}^{\eta}(x) 
V_{\nu}^{\eta}(y) V_{\rho}^{\eta}(0) \} |0 \rangle \,.
\end{array} \right .
\ee
Here $A_{\mu}^{\lambda}$ ($V_{\mu}^{\lambda}$) denote the usual
axial (vectorial) currents for the gluino fields, whereas 
$A_{\mu}^{\eta}= \bar \eta^a \gamma_\mu \eta^a$,
$V_{\mu}^{\eta}= \bar \eta^a (i \gamma_\mu \gamma_5) \eta^a$ denote the 
corresponding ones for the ghost field $\eta$.
Notice that, once again due to the Majorana nature of $\eta$ 
and its ``wrong'' statistics, the ghost-gluon coupling is of the form  
$i \gamma_\mu \gamma_5$, while the axial current insertion is 
now $\gamma_\mu$, and the free $\eta$ propagator is given by
\be
S_\eta (x, y) = - S_\lambda (x, y) (i \gamma_5) \, .
\ee
Eventually, we obtain the following relation between the triangle graphs, 
\be
\Gamma^\eta_{\mu\nu\rho} = - \Gamma^\lambda_{\mu\nu\rho}~,
\ee
with the relative minus stemming from the consideration 
of a fermion-statistics loop versus a boson-statistics one. 
It follows that the $J^0_{\mu}$ current is non-anomalous, 
as the fermion and ghost loops cancel each other exactly, 
while $J^3_{\mu}$ is anomalous as they add up. 
Hence, the chiral $U(1 \mid 1)$ group is explicitly broken 
by the $J_\mu^3$ anomalous current to a smaller group, $SU(1 \mid 1)$.
The subgroup $SU(1 \mid 1)$ admits a graded algebra with generators 
$\left\{\sigma_0, \sigma_1, \sigma_2 \right\}$.

As for the trace anomaly, it can be shown that the ghost contribution
to $\Theta^\mu_\mu$ exactly cancels the contribution of the gluino loop.
In conclusion, the anomalous chiral and scale Ward identities of the 
quenched SYM theory are
\be
\label{z2anomalies}
\left . \begin{array}{ccc}
\partial^\mu J^3_\mu &=& - (1+1) c(g) F^a_{\mu\nu} 
\tilde F^{a\mu\nu}\, , \\
\Theta^\mu_\mu &=& \frac{\beta^\prime(g)}{\beta(g)} 
c(g) F^a_{\mu\nu} F^{a\mu\nu} \, ,
\end{array} \right .
\ee
where $\beta^\prime$ is the one-loop $\beta$-function for the pure 
gauge theory, as the fermion and ghost contributions cancel out.

\section{Low energy effective theory}
\label{qVYsec}

The low-energy effective lagrangian should be invariant under the symmetries of
the fundamental theory and reproduce at the classical level 
the corresponding  anomalies, eq.~(\ref{z2anomalies}). 
In the unquenched theory it depended on two free parameters $\mu$ and $\alpha$,
as reviewed in Sect.~\ref{symsec}. For the quenched case, supersymmetry
is no longer a symmetry of the action and in principle 
many more free parameters can be introduced.
The new $U(1 \mid 1)$ symmetry provides a restrictive guideline, which we 
explore below. 

The most general effective lagrangian can be decomposed as follows:
\be
\label{mostgen}
{\cal L} = {\cal L}_{kin} + {\cal L}_{int} + {\cal L}_{anom}
\ee
The kinetic term, ${\cal L}_{kin}$, and the interaction term,
 ${\cal L}_{int}$, 
are classically invariant under chiral $U(1 \mid 1)$ symmetry and 
naive scale transformations, whereas 
\be
{\cal L}_{anom} \rightarrow {\cal L}_{anom} + {\rm trace \ \ anomaly}
\ee
under a scale transformation and
\be
{\cal L}_{anom} \rightarrow {\cal L}_{anom} + {\rm chiral \ \ anomaly}
\ee
under an anomalous chiral $U(1 \mid 1)$ transformation, 
as dictated by the structure of the fundamental theory. 
The coefficients of the two extra terms  that appear under these 
transformations are fixed by the anomalies in the fundamental theory,
eq.~(\ref{z2anomalies}).

In order to implement the $U(1\mid 1)$ symmetry, we define the following 
composite fields:
\be
\label{newfields}
\left . \begin{array}{ccc}
\hat \phi^i &=& c(g) \bar Q^a_R \sigma^i Q^a_L \, , \\
\\
\hat \chi &=& \frac{i c(g) }{2} \sigma_{\mu\nu} F^{a\mu\nu} Q^a \, .
\end{array} \right .
\ee
$\hat \phi = \sigma^i \hat \phi^i $ is a {\it scalar}
field in the adjoint representation of $U(1 \mid 1)$, transforming as
\be
\hat \phi \rightarrow  U \hat \phi U \, ,
\label{URL}
\ee
with $U$ defined in eq.~(\ref{U_11trans}).
$\hat \chi$ is a doublet {\it fermion} field\footnote{The italic characters 
{\it fermion} and {\it scalar} refer to the fact that, although $\hat \phi$
and $\hat \chi$ are a scalar and a fermion field respectively under 
Lorentz transformations, both contain component fields of mixed spin-statistics.}
defined in the fundamental representation of $U(1\mid 1)$, just like $Q^a$.

The anomalous $U_{\sigma_3}(1 \mid 1)$ transformation breaks the 
$U(1 \mid 1)$ invariance of the theory in the following way,
\be
U(1 \mid 1) \rightarrow Z_{4 N_c} \times SU(1 \mid 1)~.
\ee
The discrete $Z_{4 N_c}$ is the residual symmetry related to the 
anomalous $\sigma_3$ generator, whilst the continuous $SU(1 \mid 1)$ group 
is related to the unbroken graded subalgebra formed
by the $\{ \sigma_0, \sigma_1, \sigma_2 \}$ generators, 
as explained in Sect.~\ref{qsymsec}.
Under the assumption that the $SU(1 \mid 1)$ chiral symmetry is
spontaneously broken\footnote{This ansatz is suggested by a Coleman-Witten 
argument \cite{cw}, and supported by numerical results for quenched 
simulations \cite{dghv}.}, an exponential representation for
 the {\it scalar} field is convenient,
\be
\hat \phi = \frac{1}{\sqrt{2}}\rho \hat \Sigma = \frac{1}{\sqrt{2}} 
\rho e^{ i \hat \theta } \, , 
\label{exprepr}
\ee
with $\rho$ a {\it scalar} field invariant under 
$U(1 \mid 1)$ and $\hat \Sigma$ a {\it pseudoscalar} field  
($\hat \theta = \hat \theta^i \sigma^i, i = 0,\dots,3$ ).
The three {\it pseudoscalars}  $\theta^i$ ($i = 0,1,2$) 
are the would-be Goldstone modes associated with the spontaneous 
breaking of $SU(1 \mid 1)$. The pseudoscalar associated
to the anomalous generator is $ \theta_3 = -\frac{i}{2} Str \log \Sigma$.

Consider now ${\cal L}_{anom}$. Under a $\sigma_3$ transformation, 
\be
\label{t3trans}
\theta_3 \to \theta_3 + \alpha_3.
\ee
There is only one dimensionful field, $\rho$. Under a scale transformation, 
\be
\label{rhotrans}
\rho \to e^\gamma \rho .
\ee
Finally, both $\theta_3$ and $\rho$ are invariant under $SU(1 \mid 1)$. 
Hence, the anomalous term in the lagrangian, as fixed by eq. (\ref{z2anomalies}), 
is
\be
{\cal L}_{anom} = - (1+1) i ( M - M^\dagger) \theta_3 
- \frac{\beta^\prime}{\beta} ( M + M^\dagger) 
\log \frac{\rho}{\bar \mu}\,, 
\label{Lanom}
\ee
with
\be
\left . \begin{array}{ccc}
M - M^\dagger &=& - i c(g) F \tilde F \,, \\
\\
M + M^\dagger &=& - c(g) F F \,. 
\end{array} \right .
\ee

Let us turn now to the most general form for ${\cal L}_{kin}$,
\bea
{\cal L}_{kin} &=& 
V_1(\rho) \left ( \frac{1}{2} \partial_\mu \rho \partial^\mu \rho \right ) +
V_2(\rho) \left [
\frac{\rho^2}{2} Str \left ( \partial_\mu \hat \Sigma^\dagger 
                             \partial^\mu \hat \Sigma \right ) 
\right ] \nn \\
&+& V_3(\rho) \left ( 
i \bar{\hat \chi}_R \gamma^\mu \partial_\mu \hat \chi_R \right ) +
V_4(\rho) \rho^2 \partial_\mu \theta_3 \partial^\mu \theta_3\,.
\label{lagkin}
\eea
Notice that we have written the coefficients $V_i$ as arbitrary functions 
of the invariant field $\rho$:
\begin{itemize}
\item
They cannot depend on $\theta_3$ and/or $\log(\rho/\bar \mu)$, as this
would spoil the desired anomalies by introducing supplementary contributions
in addition to the correct ones generated by the term ${\cal L}_{anom}$ above.
\item
They could depend on $\hat \Sigma$ in a $SU(1 \mid 1)$ invariant form, like
$Str ( \partial_\mu^n \hat \Sigma^\dagger \partial^{n\mu} \hat \Sigma )$.
However, since we consider the leading order in the momentum expansion, no new
derivatives should be added in the coefficients in front of the kinetic terms.
We are thus left with $Str( \hat \Sigma^\dagger \hat \Sigma) = 0$, and as a result
the coefficients $V_i$ do not depend on $\hat \Sigma$, either.
\end{itemize}

Consider now the interaction lagrangian ${\cal L}_{int}$. 
The most general expression invariant under $U(1 \mid 1)$, 
at zero order in the derivative expansion, is given by
\bea
{\cal L}_{int} &=& \sum_{m=1}^\infty a_m(\rho) \rho^m  
+ \sum_{m,n,p=0}^\infty 
\left ( \rho \bar{\hat \chi}_R \hat \Sigma \hat \chi_L \right )^m 
\left ( \rho \bar{\hat \chi}_L \hat \Sigma^\dagger \hat \chi_R \right )^n 
\nn \\
&\times& \left [ c_{mnp}(\rho) (M - M^\dagger)^p + d_{mnp}(\rho) (M + M^\dagger)^p 
\right ] 
\label{Lint}
\eea
where $a_m, c_{mnp}$ and $d_{mnp}$ parametrize all the unknown 
effects of the quenching. Their field dependence only encompasses 
the field $\rho$, as the same arguments used above with respect 
to the coefficients $V_i$ in ${\cal L}_{kin}$ hold here.

A supplementary dimensional argument should be applied both to
${\cal L}_{kin}$ and  ${\cal L}_{int}$, in order to further constraint
their parameters.  Since in the fundamental theory there are no 
explicit mass parameter, the effective theory must be scale invariant up
to the trace anomaly (the same analysis has been done for QCD in \cite{gjjs}).
No terms with a dimensionful parameter can be introduced
in the lagrangian. As a consequence, the correct dimensionality of any
term can be fulfilled by inserting the appropriate powers of $\rho$.
In what concerns  ${\cal L}_{kin}$, this means that the coefficients 
$V_i(\rho)$ of the kinetic lagrangian are just constants, $V_i$. 
Moreover, we can explicitly write the $\rho$-dependence in ${\cal L}_{int}$:
\bea
{\cal L}_{int} &=& \sum_{m=1}^\infty a_m \rho^m 
+ \sum_{m,n,p=0}^\infty 
\frac{1}{\rho^{4 (m+n+p) - 4} }
\left ( \rho \bar{\hat \chi}_R \hat \Sigma \hat \chi_L \right )^m 
\left ( \rho \bar{\hat \chi}_L \hat \Sigma^\dagger \hat \chi_R \right )^n \nn \\
&\times& \left [ c_{mnp} (M - M^\dagger)^p + d_{mnp} (M + M^\dagger)^p 
\right ]
\label{Lint2}
\eea
where the coefficients $a_m, c_{mnp},d_{mnp}$ are constants,
the original coefficients having been 
redefined so as not to depend on the fields anymore. 

This latter analysis deserves a comment: in principle only an analytical 
dependence on $\rho$ can be removed by dimensional arguments. 
Nevertheless, we consider a reasonable assumption that the only possible 
non-analytical field dependence is in ${\cal L}_{anom}$:
the anomalies are manifestly non-analytical in the ghost field, as they vary
discontinously when the ghost field is turned on, eq. (\ref{z2anomalies}).

Gathering all the above results, a final step is now pertinent.
The fundamental action of the unquenched SYM should be, and is, smoothly
recovered from the quenched SYM theory when the ghost field is switched off,
$\eta \to 0$, except for the anomalous non-analytical contributions.
Analogously, we expect to recover $S_{VY}$ from our effective action when
 all ghost-like fields are removed, except again for the anomalous 
contributions, which should remain the same. This assumption further constraints
the coefficients of the effective lagrangian. In order to study
this limit, develope first the kinetic lagrangian in a canonical form:
\bea
\label{lagkin2}
{\cal L}_{kin} &=& 
\frac{1}{2} \partial_\mu \rho \partial^\mu \rho +
\frac{\rho^2}{2} \left ( \partial_\mu \theta \partial^\mu \theta - 
\partial_\mu \tilde \theta \partial^\mu \tilde \theta \right ) \nn \\
&+& \rho^2 \partial_\mu \theta^+ \partial^\mu \theta^- 
+ \left ( i \bar{\hat \chi}_R \gamma^\mu \partial_\mu \hat \chi_R \right ) \nn \\
&+& \left ( \frac{V_4}{V_1} \right )
\rho^2 \partial_\mu \theta_3 \partial^\mu \theta_3 \,,
\eea
where
\be
\left . \begin{array}{ccc}
\theta_0 &=& \frac{1}{2} ( \theta + \tilde \theta ) \,, \\
\\
\theta_3 &=& \frac{1}{2} ( \theta - \tilde \theta ) \,.
\end{array} \right .
\ee
$\theta$ and $\tilde \theta$ behave like $\bar \lambda^a \gamma_5 \lambda^a$ 
and $\bar \eta^a \gamma_5 \eta^a$
respectively. In eq. (\ref{lagkin2}) the following rescaling has been 
performed:
\be
\rho \to \frac{\rho}{\sqrt{V_1}} ; \qquad \theta_i \to \theta_i \sqrt{\frac{V_1}{V_2}} ;
\qquad \chi \to \frac{\chi}{\sqrt{V_3}}\,. \nn
\ee
It can be noticed that the kinetic term of $\hat \Sigma$ contains fields 
with a ``wrong'' metric, a left-over of their ghostness.
In the limit $\eta \to 0$, we get:
\bea
\rho &\to& \rho_S \,,\nn \\
\theta &\to& \theta_S \,, \nn \\
\tilde \theta &\to& 0  \,, \\
\theta^\pm &\to& 0 \nn \,, \\
\hat \chi &\to& \left ( \begin{array}{c} \chi_S \\ 0 \end{array} 
\right) \,,\nn
\eea
where the subscript $S$ refers to the fields in the supersymmetric 
$S_{VY}$ action, eq. (\ref{offVYcan}).

Matching, thus, the VY action and the quenched action in the $\eta \to 0$
limit, fixes uniquely the coefficients of the latter. It follows:
\bea
S^q_{VY} &=& \int d^4 x \left \{ 
  \frac{1}{2} \partial_\mu \rho \partial^{\mu} \rho +
  \frac{\rho^2}{2} 
      Str \left ( \partial_\mu \hat \Sigma^\dagger \partial^\mu \hat \Sigma \right ) +
 i \bar{\hat \chi}_R \gamma^\mu \partial_\mu \hat \chi_R
\right . \nn \\
&+& \left . 
2 \frac{ \left( \bar{\hat \chi}_R \hat \Sigma \hat \chi_L\right) 
         \left( \bar{\hat \chi}_L \hat \Sigma^{\dag} \hat \chi_R \right)}{\rho^2}
- \frac{9}{\alpha^{3/2}} \frac{2 \sqrt{2} }{\rho^3}  
\left( M \bar{\hat \chi}_L \hat \Sigma^{\dag} \hat \chi_R +
       M^\dagger \bar{\hat \chi}_R \hat \Sigma  \hat \chi_L \right) 
\right . \nn \\
&+& \left . 
\frac{81}{\alpha^3 } \frac{4}{\rho^4} M^\dagger M
- \frac{\alpha^{3/2}}{18} \frac{\rho}{\sqrt{2}} 
\left( \bar{\hat \chi}_L \hat \Sigma^{\dag} \hat \chi_R
     + \bar{\hat \chi}_R \hat \Sigma \hat \chi_L \right) 
\right . \nn \\
&-& \left . 
Str \left [ \hat J_\mu \hat \Sigma \dedouble \hat \Sigma^\dagger \right ] \right . \nn \\
&-& \left . 2 (1+1) i ( M - M^\dagger) \theta_3 
- \frac{\beta^\prime}{\beta} ( M + M^\dagger) \log \frac{\rho}{\bar \mu} 
\right \} .
\label{final}
\eea
where $\hat J_\mu = \sigma^i 
( \bar{\hat \chi}_R \sigma^i \gamma_\mu \hat \chi_R )$\footnote{
We have verified that this term reduces to the corresponding one in the VY lagrangian
in the $\eta \to 0$ limit, although we have not studied other possible allowed contributions
for this particular term. Nevertheless, its functional form is not relevant for the mass
spectrum of the theory, our principal interest in this paper.}. 
This new action is no longer supersymmetric, although it strongly
resembles eq. (\ref{offVYcan}). 

The equation of motion for the auxiliary field is:
\be
M = \left ( \frac{\alpha^3}{81} \frac{\rho^4}{4} \right ) 
\left [ - (1+1) 2 i \theta_3 
+ \left ( \frac{\beta^\prime}{\beta} \right ) \log \frac{\rho}{\bar \mu} +
\frac{9}{\alpha^{3/2}} \frac{2 \sqrt{2} }{\rho^3} \bar{\hat \chi}_R \hat \Sigma  \hat \chi_L  
\right ]\,,
\ee 
and the on-shell action is:
\bea
S^q_{VY} &=& \int d^4 x \left \{ 
  \frac{1}{2} \partial_\mu \rho \partial^{\mu} \rho +
  \frac{\rho^2}{2} 
      Str \left ( \partial_\mu \hat \Sigma^\dagger \partial^\mu \hat \Sigma \right ) +
 i  \bar{\hat\chi}_R \gamma^\mu \partial_\mu \hat \chi_R
\right . \nn \\
&-& \left . 
\frac{\alpha^{3/2}}{9} \frac{\rho}{\sqrt{2}}
\left [ \frac{\beta^\prime}{\beta} \log \frac{\rho}{\bar \mu} 
+ 2 ( 1+1) i \theta_3 \right ]
\bar{\hat \chi}_R \hat \Sigma  \hat \chi_L \right . \nn \\
&-& \left . 
\frac{\alpha^{3/2}}{9} \frac{\rho}{\sqrt{2}}
\left [ \frac{\beta^\prime}{\beta} \log \frac{\rho}{\bar \mu} 
- 2 ( 1+1) i \theta_3 \right ]
\bar{\hat \chi}_L \hat \Sigma^{\dag} \hat \chi_R \right . \nn \\
&-& \left . 
\frac{\alpha^3}{81} \frac{\rho^4}{4} 
\left [ \left (\frac{\beta^\prime}{\beta}\right)^2 \log^2 \frac{\rho}{\bar \mu}
+ 4 (1+1)^2 \theta_3^2 \right ] \right . \nn \\
&-& \left .  \frac{\alpha^{3/2}}{18} \frac{\rho}{\sqrt{2}} 
\left( \bar{\hat \chi}_L \hat \Sigma^{\dag} \hat \chi_R
     + \bar{\hat \chi}_R \hat \Sigma \hat \chi_L \right) 
\right . \nn \\
&-& \left . 
Str \left [ \hat J_\mu \hat \Sigma \dedouble \hat \Sigma^\dagger \right ]
\right \} \, .
\label{final_on}
\eea
It is interesting to notice that eq.~(\ref{Lint}) 
could have been greatly simplified, would have we considered
the leading order in the $1/N_c$ expansion. 
In the Appendix we show that only a small subset of parameters
must be fixed then by requiring that $S_{VY}$ is recovered in the $\eta \to 0$ limit. 

\section{The mass spectrum}
\label{secmass}

The aim of this paper is to give a prediction for the spectrum
of the quenched SYM theory, and in particular for the 
mass splitting induced by the quenched approximation.
In the previous section, the scalar potential was found to be:
\be
\label{vvv}
V(\rho,\theta_3) = 
\frac{\alpha^3}{81} \frac{\rho^4}{4}
\left[ 4 (1+1)^2 \theta_3^2 
+ \left (\frac{\beta^\prime}{\beta} \right )^2  
\log^2 (\frac{\sqrt{\alpha}}{3}\frac{\rho}{ \sqrt{2} \mu' }) \right]
\ee
where we have redefined $\bar \mu =\frac{3 \sqrt{2} }{\sqrt{\alpha}} \mu'$ so as to
use a notation alike to that in ref.\cite{vy}, the {\it prime} recalling
that, in our knowledge, there is no reason for the nonperturbative 
mass scale of the quenched theory to be the same as the unquenched one.
 
The mass spectrum is derived by expanding eq.~(\ref{final_on}) around the 
minimum of $V$, eq. (\ref{vvv}),
located at $\langle 0|\rho|0 \rangle=\bar\mu$, 
and $\langle 0|\theta_3|0 \rangle=0$. 
Shifting the field $\rho  \to \bar \mu + \sigma$, 
rescaling $\theta^i \to \theta^i / \bar{\mu}$ and
developing $\theta_{0,3}$ in terms of  $\theta$ and $\tilde{\theta}$,
it follows:
\bea
{\cal L}  &=& \frac{1}{2} \partial^\mu \sigma \partial_\mu \sigma 
+ \frac{1}{2}\left [ \partial^\mu \theta \partial_\mu \theta -
\partial^\mu \tilde \theta \partial_\mu \tilde \theta \right ] 
+ \partial_\mu \theta^+ \partial^\mu \theta^- 
+ i \bar{\hat \chi}_R \gamma^\mu \partial_\mu \hat \chi_R \nn \\ 
&-& (1+1)^2 \frac{m^2_{\chi}}{2}
\left(\theta-\tilde{\theta}\right)^2
-\left(\frac{\beta^{\prime}}{\beta}\right)^2\frac{m^2_{\chi}}{2}
\sigma^2
- \frac{m_{\chi}}{2}\bar{\hat \chi} \hat \chi \, ,
\label{Lfree}
\eea
where $m_{\chi}=\alpha \mu' / 3$ and
$\theta^{\pm}=(\theta_1 \pm i\theta_2)/\sqrt{2}$ are two 
massless {\it pseudoscalar} Goldstone fermions.
As for $\theta$ and $\tilde{\theta}$, their non-diagonal
mass matrix implies that their would-be mass term cannot be resummed: 
the mass term is to be taken as an interaction vertex, alike to the
situation in \cite{bg}. The Euclidean propagator is then:
\bea
G_\theta (p) &=& \frac{1}{p^2} \left [ 1 - \frac{(1+1)^2 m_\chi^2}{p^2} \right ] 
\nn \\
\\
G_{\tilde \theta} (p) &=& - \frac{1}{p^2} \left [ 1 + \frac{(1+1)^2 m_\chi^2}{p^2} \right ]
\nn
\eea
(the graphs with multiple $\theta^2, \tilde \theta^2$ insertions 
cancel out those with $\theta \tilde \theta$ insertions, 
and as a consequence the single insertions result in a double-pole propagator). 

The mass spectrum for the bound states of the gluino and gluon
fields is then given by
\bea
m_\sigma &=& \frac{\beta^\prime}{\beta} m_\chi \, ,\nn \\
m_\chi   &=& \frac{1}{3} \alpha \mu', \label{massspec} \\
m_\theta &=&  (1+1) m_\chi ,\nn 
\eea
to be compared with $m_\chi \,=\, m_\sigma \,=\, m_\theta$ in the unquenched
theory. It is not completely correct to consider the coefficient $(1+1)^2 m_\chi^2$
of the double-pole term in $G_\theta(p)$ as a mass term for the $\theta$ pseudoscalar,
since it is not possible to resum it, as explained above. 
However, in a numerical simulation that coefficient
is interpreted as its would-be mass (to become a proper mass term after resummation of
the internal fermion loops in the full theory).

The spectrum for the scalar ($\rho$) and fermion ($\chi$) 
fields shows that the mass splitting of the VY supermultiplet 
results from the non-analiticity of the anomaly dependence  
on the ghost field.

In ref.~\cite{dghv} results for the quenched mass spectrum 
of $N=1$ SYM with gauge group $SU(2)$ 
were presented, with $m_\chi = 0.58(9)$ and $m_\sigma = 0.64(6)$
in the chiral limit. The ratio $(m_\sigma/m_\chi)_{lat} = 1.1(3)$ 
is in fair agreement with our theoretical expectation in eq. (\ref{massspec}), 
$(m_\sigma/m_\chi)_{th} = 11/9 = 1.22$ for $SU(2)$. 
However, further quenched simulations are needed 
in order to decrease the numerical errors and to confirm 
the theoretical prediction for the $\sigma-\chi$ splitting, 
as well to study the $\theta-\chi$ splitting in full detail.

\section{Conclusions}
\label{concl}
 
We have implemented quenching on $N=1$ SYM theory, in the continuum,
by introducing a ghost-like field, which cancels the internal fermion-loop
effects on observable quantities.
We have then derived the corresponding low energy effective lagrangian, the 
relevant object with respect to the spectrum and interactions of the bound states 
of the theory.

Although supersymmetry is lost upon quenching, it turns out that a new
$U(1 \mid 1)$ symmetry arises, explicitly broken by the chiral anomaly
to $Z_{4 N_c} \times SU(1 \mid 1)$. 
Its rich and beautiful anomaly structure entails a controlable splitting 
of the Veneziano-Yankielowicz multiplet: 
the pseudoscalar mass doubles the fermionic one, and the scalar mass is 
about 20 \% heavier than the latter (for the $SU(2)$ gauge group).
   
These results provide a first estimate of the systematic error associated
to quenching in lattice supersymmetry computations, and illuminate recent 
numerical computations.

\section*{Acknowledgements.}

We thank A. Gonzalez-Arroyo, P. Hernandez, G. C. Rossi, M. Testa
and A. Vladikas for useful discussions. 

\section*{Appendix: $1/N_c$ expansion}
\label{app}

The interaction lagrangian ${\cal L}_{int}$ of eq.~(\ref{Lint}) can be strongly simplified
by means of $1/N_c$ arguments, in the spirit of ref.~\cite{dvv}.

${\cal L}_{int}$ represents the interaction potential of the fundamental theory. 
At the fundamental level, the effective potential is the resummation of the 1PI 
graphs of the theory. Let us first recall the situation in QCD. There,
the potential can be organized in powers of fermion loops by using the 
$1/N_c$ expansion, since the leading term in $1/ N_c$ for fermions in the 
fundamental representation (the quarks) is of ${\cal O}(N_c)$, 
whereas any internal fermion loop is suppressed as ${\cal O}(1/ N_c)$ 
with respect to an internal gluon loop. 
The leading term of this expansion is then represented by one fermion loop plus
the resummation of all the internal gluon loops; the subleading ${\cal O}(1 /N_c)$ term  
contains two fermion loops plus internal gluon loops and so on. 
For this reason, the $1/N_c$ expansion and the quenched
approximation are intimately related for QCD, as quenching can be regarded
to be the leading term of $1/N_c$ expansion at fixed $N_f$. 
 
In SYM, instead, bosonic and fermionic loops are of the same order in $1/N_c$, ${\cal O}(N_c^2)$,
since the fermions are in the same representation as the gluons. It is not possible, hence, 
to separate gluons and gluino loops, and ${\cal L}_{int}$ is ${\cal O}(N_c^2)$. 
However, recalling that $(M \pm M^\dagger)$ and $\rho$ are ${\cal O}(N_c)$ 
(when correctly normalized, see again \cite{dvv}), we can extract the $N_c$ dependence 
of the coefficients:
\be
\left . \begin{array}{ccc}
a_m & = & {\cal O} \left ( \frac{1}{N_c^{m - 2} } \right ) \, , \\
\\
c_{mnp},d_{mnp} & = & {\cal O} \left ( \frac{1}{N_c^{m + n + p - 2 }} \right ) \, .
\end{array} \right .
\ee

Hence, at the leading order in the $1/N_c$ expansion, we get:
\bea
{\cal L}_{int} &=& \sum_{m,n,p=0}^2  
\left \{ \frac{1}{\rho^{4 (m+n+p) - 4} }
\left ( \rho \bar{\hat \chi}_R \hat \Sigma \hat \chi_L \right )^m 
\left ( \rho \bar{\hat \chi}_L \hat \Sigma^\dagger \hat \chi_R \right )^n 
\right . \nn \\
&\times& \left .
\left [ c_{mnp} (M - M^\dagger)^p + d_{mnp} (M + M^\dagger)^p \right ]
\right \} \nonumber \\
&+& {\cal O } \left ( \frac{1}{N_c} \right ) \, , 
\label{Lint_1/N}
\eea
where $m+n+p \le 2$. We have used dimensional arguments 
to extract the $\rho$-dependence
from the coefficients $c_{mnp}, d_{mnp}$ and $a_m$ (the only dimensionally
allowed term of this type would be $a_4$, suppressed as $1/N_c^2$).

Expanding this lagrangian, we get:
\bea
{\cal L}_{int} &=& 
(c_{100} + d_{100}) 
\left ( \rho \bar{\hat \chi}_R \hat \Sigma \hat \chi_L \right ) +
(c_{010} + d_{010}) 
\left ( \rho \bar{\hat \chi}_L \hat \Sigma^\dagger \hat \chi_R \right ) 
\nn \\
&+& 
(c_{200} + d_{200}) 
\frac{ \left ( \rho \bar{\hat \chi}_R \hat \Sigma \hat \chi_L \right )^2}{\rho^4} +
(c_{020} + d_{020}) 
\frac{ \left ( \rho \bar{\hat \chi}_L \hat \Sigma^\dagger \hat \chi_R \right )^2}{\rho^4} 
\nn \\
&+&
(c_{110} + d_{110}) 
\frac{ \left ( \rho \bar{\hat \chi}_R \hat \Sigma \hat \chi_L \right )
\left ( \rho \bar{\hat \chi}_L \hat \Sigma^\dagger \hat \chi_R \right ) }{\rho^4}
\nn \\
&+& c_{001} (M - M^\dagger) + d_{001} (M + M^\dagger) \nn \\
&+& (c_{002} + d_{002}) \frac{ ( M^2 + M^{\star 2} )}{\rho^4} 
- 2 (c_{002} - d_{002}) \frac{ M^\dagger M }{\rho^4} \nn \\
&+&
\left [ 
(c_{101} + d_{101}) \left ( \rho \bar{\hat \chi}_R \hat \Sigma \hat \chi_L \right ) +
(c_{011} + d_{011}) \left ( \rho \bar{\hat \chi}_L \hat \Sigma^\dagger \hat \chi_R \right ) 
\right ] \frac{M}{\rho^4} \nn \\
&-&
\left [ 
(c_{101} - d_{101}) \left ( \rho \bar{\hat \chi}_R \hat \Sigma \hat \chi_L \right ) +
(c_{011} - d_{011}) \left ( \rho \bar{\hat \chi}_L \hat \Sigma^\dagger \hat \chi_R \right ) 
\right ] \frac{M^\dagger}{\rho^4} \nn \\
&+& {\cal O } \left ( \frac{1}{N_c} \right ) 
\label{Lint_1/N_2}
\eea
Terms linear in $(M \pm M^\dagger)$ are $\theta$-angles. 
Only a small set of free parameters remains at the leading order in
$1/N_c$, then, to be 
fixed requiring the recovery of the VY lagrangian in the $\eta \to 0$ limit. 
Our final results for the mass spectrum are, of course, not modified by 
the results of this Appendix.

\end{document}